\documentclass[useAMS,usenatbib]{mn2e}\voffset=-1cm

\usepackage{epsfig}
\usepackage{rotating}
\usepackage{xcolor}
\usepackage{graphicx}
\usepackage{epstopdf}
\usepackage{verbatim}

\newif\ifAMStwofonts
\AMStwofontstrue

\title[The {\it Herschel}-SPIRE Dark Field]{The {\it Herschel}-SPIRE Dark Field I: The deepest {\it Herschel} image of the submillimetre Universe}
\author[C.~Pearson  {\it et al.}]
       {
       Chris Pearson$^1$$^{,2}$$^{,3}$\thanks{E-mail: chris.pearson@stfc.ac.uk},
       Thomas W. O. Varnish$^{4,1,5}$,
       Xinni Wu$^4$,
       David L. Clements$^4$,
       \vspace*{0.3cm}\\ 
        {\LARGE  \textup{
       Ayushi Parmar$^4$,
       Helen Davidge$^2$,
       Matthew Pearson$^1$$^{,6}$
       }}
\vspace*{0.3cm}\\ 
\vspace*{0.1cm}\\
      $^1$ RAL Space, UKRI STFC Rutherford Appleton Laboratory, Chilton, Didcot, Oxfordshire OX11 0QX, UK\\
      $^2$ Department of Physical Sciences, The Open University, Milton Keynes, MK7 6AA, UK\\
      $^3$ Oxford Astrophysics, Denys Wilkinson Building, University of Oxford, Keble Rd, Oxford OX1 3RH, UK\\
      $^4$ Imperial College London, Blackett Laboratory, Prince Consort Road, London, SW7 2AZ, UK\\
      $^5$ Plasma Science and Fusion Center, Massachusetts Institute of Technology, Cambridge, MA 02139, USA\\
      $^6$ Catalent Pharma Solutions, Frankland Rd, Swindon SN5 8YG, UK}
\date{Accepted .\\
      Received ;\\
      in original form 2023 April }

\pagerange{336--346}

\begin{document}

\label{firstpage}

\maketitle

\begin{abstract}
We present the image maps, data reduction, analysis and the first source counts from the {\it Herschel} SPIRE Dark Field. The SPIRE Dark Field is an area of sky near the North Ecliptic Pole observed many times during the calibration phase of the {\it Herschel} mission in order to characterise the stability of the SPIRE instrument and is subsequently one of the deepest imaged fields of the Universe at far-infrared-submillimetre wavelengths. The SPIRE dark field is concurrent with the {\it Spitzer} IRAC Dark Field used for a similar purpose. The final Dark Field map is comprised of 141 individual SPIRE observations in Small Map and Large Map modes defined by a deep inner region approximately 12$\arcmin$ in diameter and a slightly shallower surrounding area of diameter $\sim$30$\arcmin$. The depth of both regions reach well below the confusion limit of the SPIRE instrument at 250$\, \umu$m, 350$\, \umu$m and 500$\, \umu$m.
Two independent processes are used to extract sources, a standard map based method using the SUSSEXtractor algorithm and a list driven photometry approach using the XID algorithm with the  {\it Spitzer} MIPS 24$\, \umu$m catalogue as an input prior.  The resulting source counts detect the turnover in the galaxy population with both methods shown to be consistent with previous results from other {\it Herschel}  surveys, with the XID process reaching approximately twice as deep compared to traditional map based algorithms. Finally, we compare our results with two contemporary galaxy evolution models, again showing a good general agreement with the modelled counts.
\end{abstract}



\section{Introduction}\label{sec:introduction}
The cosmic infrared background contains the integrated history of obscured star-formation in the Universe showing that half of the energy output throughout cosmic history has been emitted long-ward of $\sim$1$\, \umu$m \citep{dole06}. Clearly, observations at far-infrared to sub-millimetre wavelengths are vital to understand how stars and galaxies have evolved over cosmic time. Atmospheric absorption necessitates that these observations require space-borne facilities, the most recent being the {\it Herschel} space observatory \citep{pilbratt10}. {\it Herschel}  observed in the far-infrared in 3 wavelength bands with the PACS instrument (70, 100, 160$\, \umu$m, \citealt{poglitsch10}) and 3 wavelength bands in the sub-millimetre with the SPIRE instrument  (250, 350, 500$\, \umu$m, \citealt{griffin10}). Large scale galaxy surveys were carried out as part of both the guaranteed time observations (the {\it Herschel} Multi-tier Extragalactic Survey, HerMES, \citealt{oliver12}) and the key open time observations (H-ATLAS, \citealt{eales10}).  The HerMES programme followed a Wedding-Cake hierarchical structure covering wide shallow fields of 100's square degrees e.g. HeLMS \citep{asboth16}, fields of the order 5-10 square degrees such as COSMOS \citep{scoville07}), to deep, narrow fields of $\sim$0.5 square degree covering the GOODS fields \citep{giavalisco04}.

These surveys cover a range in flux density from 1~Jy down to 10s of mJy and have provided an excellent constraint for models of galaxy evolution. The {\it Herschel} surveys require strong evolution in the infrared galaxy population in order to produce the steep rise in the galaxy source counts below 100~mJy with all pre-{\it Herschel} models struggling to match the observations (\citealt{lagache04}, \citealt{negrello07},  \citealt{rowanrobinson09}, \citealt{pearson09}, \citealt{valiante09}).

The deepest fields observed with the SPIRE instrument also suffer from acute confusion due to the crowded fields combined with large instrument  beam sizes (full width half maximum, FWHM = 17.6, 23.9, 35.2$\arcsec$ in the PSW (250$\, \umu$m), PMW (350$\, \umu$m) and PLW (500$\, \umu$m) bands respectively). This makes both the source detection and corresponding association with multi-wavelength data (e.g. optical sources) challenging. Source blending has a major effect on identifications with the effect becoming more critical at longer wavelengths and deeper flux densities \citep{scudder16}.

 In this paper we present a new extremely deep {\it Herschel} field comprised of a large number of observations taken during the calibration time of the SPIRE instrument of the {\it Herschel} mission. The SPIRE dark field is one of the deepest sub-millimetre fields on the entire sky. In Section \ref{sec:Observations}, we describe the  {\it Herschel} SPIRE observations. In Section  \ref{sec:Reduction} we describe the data reduction steps using standard processing pipelines. In Section \ref{sec:SourceExtraction}, we describe the source extraction process and photometry measurements using a standard map-based extraction technique, with the corresponding galaxy source counts presented in Section~\ref{sec:SourceCounts}. Due to the confused nature of the SPIRE Dark Field, in Section~\ref{sec:xid} we describe an alternative list-driven process for our source extraction and photometry. Our results are presented in Section \ref{sec:Results} where we compare the results of the Dark Field observations with other SPIRE surveys and also with contemporary galaxy evolution models. Our conclusions and summary are given in Section \ref{sec:Conclusions}. Throughout this work we assume a Hubble constant of  $H_0=70$\,km\,s$^{-1}$\,Mpc$^{-1}$ and density parameters of $\Omega_{\rm M}=0.3$ and $\Omega_\Lambda=0.7$.


\section{Observations}\label{sec:Observations}
The SPIRE Dark Field is an area of sky, dedicated to the calibration of {\it Herschel}. Centred near but approximately  3.5 degrees away from the North Ecliptic Pole at R.A. = 17h 40m 12s, Dec = +69d 00m 00s, the field is devoid of any bright sources and has relatively low background emission $\sim$2~MJy/sr at 857 GHz (350$\, \umu$m) as measured by the Planck\footnote{ \tt {https://irsa.ipac.caltech.edu/data/Planck/} \\  \tt {release$\_$3/all-sky-maps/}} satellite \citep{planck11}. The field was observed extensively during the  {\it Herschel} Performance Verification (PV) phase to measure standard load curves, conduct multi-level noise tests and to monitor detector behaviour. During the operational phase of the {\it Herschel}  mission, the SPIRE dark field was also observed regularly on an approximately one-week cadence as part of the SPIRE routine calibration plan  to monitor the detector behaviour (e.g. drifts, etc) and for flux calibration measurements \citep{pearson14}. These weekly dark field observations consisted of 4 repetitions of a 30$\arcmin \times$ 30$\arcmin$ field of view  map in the SPIRE Large Scan Map mode and a further 4 repetitions of an 8$\arcmin \times$ 8$\arcmin$ field of view observation in the SPIRE Small Map mode.

The SPIRE Dark Field is partly coincident with the  similar dark field area used for the calibration of the InfraRed Array Camera (IRAC; \citealt{fazio04}) on the Spitzer Space Telescope \citep{werner04}. The IRAC Dark Field is comprised of a $\sim$20$\arcmin$ diameter area (380 arcmin$^2$) observed in all 4 IRAC bands (3.6, 4.5, 5.8,  8.0$\, \umu$m) and is described in detail in \citet{krick09, krick12}. Multi-wavelength observations were also carried out from optical to X-ray wavelengths including mid-infrared wavelengths from {\it AKARI} \citep{davidge17} and {\it Spitzer}-MIPS at  24$\, \umu$m \citep{krick09}. 

The {\it Herschel Science Archive} was searched for all SPIRE photometry observations within a 0.5 degree radius around the centre, R.A. = 17h40m12s, Dec = +69d00m00s, of the SPIRE Dark Field. In order to create a robust and reliable set of observations with similar instrument and observational parameters, the following criteria were applied to our selection process:
\begin{itemize}
\item Only observations using the SPIRE large and small map modes: {\it SpirePhotoSmallScan} or {\it SpirePhotoLargeScan} (or equivalent) were selected.
\item Only observations made after {\it Herschel} Operational Day OD99 were selected. The scan map mode for SPIRE was optimised on OD98-99. Earlier observations formed part of the SPIRE Performance Verification Phase and were thus deemed unreliable and discarded. 
\item Select only maps observed at the nominal scan rate of 30$\arcsec$ s$^{-1}$.
\item Remove all abnormally long scans where the map size is 60$\arcmin \times$ 1$\arcmin$.
\end{itemize}

These selection criteria resulted in a total of 142 individual observations, of which one further (observation ID 1342182491) was rejected since it could not be processed through the pipeline due to partial missing data. The final total of 141 observation IDs are listed in Table~\ref{tab:observations}.

\begin{table*}
\caption{The final 141 SPIRE observations in the SPIRE Dark Field region selected from the {\it Herschel Science Archive}.}
\begin{tabular}{@{}llllllllll}
 \multicolumn{9}{l}{Herschel Observation ID} \\
\hline
1342182484 & 1342182485 & 1342182486 & 1342182487 & 1342182488 & 1342182490 & 1342182494 & 1342182496 & 1342183372  \\
1342183494 & 1342183496 & 1342183498 & 1342183500 & 1342183502 & 1342183504 & 1342183505 & 1342185820 & 1342185829  \\
1342186852 & 1342186853 & 1342186854 & 1342186855 & 1342187454 & 1342187457 & 1342187458 & 1342187459 & 1342187517  \\
1342188177 & 1342188592 & 1342188827 & 1342189528 & 1342189700 & 1342190293 & 1342190667 & 1342191195 & 1342192080  \\
1342192082 & 1342193027 & 1342193028 & 1342193802 & 1342193804 & 1342195323 & 1342195324 & 1342195689 & 1342195690  \\
1342195758 & 1342195759 & 1342195983 & 1342195984 & 1342196672 & 1342196673 & 1342196905 & 1342196906 & 1342197326  \\
1342197327 & 1342197330 & 1342197331 & 1342198153 & 1342198155 & 1342198583 & 1342198584 & 1342199368 & 1342199369  \\
1342199762 & 1342199763 & 1342200190 & 1342200191 & 1342204108 & 1342204359 & 1342204941 & 1342205083 & 1342206194  \\
1342206684 & 1342207044 & 1342209319 & 1342210544 & 1342210561 & 1342210883 & 1342211401 & 1342212305 & 1342212364 \\
1342213202 & 1342213449 & 1342214567 & 1342214704 & 1342215988 & 1342216915 & 1342218686 & 1342218970 & 1342219817 \\
1342219962 & 1342220839 & 1342221470 & 1342221927 & 1342222119 & 1342222597 & 1342222835 & 1342223223 & 1342224025 \\
1342224962 & 1342227000 & 1342227741 & 1342228356 & 1342229134 & 1342230882 & 1342231330 & 1342233315 & 1342234916 \\
1342236259 & 1342237510 & 1342238341 & 1342239275 & 1342241172 & 1342244172 & 1342244839 & 1342245159 & 1342245420 \\
1342245576 & 1342245901 & 1342246575 & 1342247283 & 1342247979 & 1342248490 & 1342249112 & 1342249228 & 1342250629 \\
1342250807 & 1342253439 & 1342254099 & 1342254480 & 1342255052 & 1342255088 & 1342256902 & 1342257357 & 1342259481 \\
1342261595 & 1342263864 & 1342265406 & 1342267753 & 1342268366 & 1342270214 & & & \\
\hline
\end{tabular}
\label{tab:observations}
\end{table*}


\section{Data Reduction}\label{sec:Reduction}

All {\it Herschel} SPIRE Data was  processed using the Herschel Common Science System  {\it Herschel Interactive Processing Environment} (HIPE, \citealt{ott10}) through the standard SPIRE Large Map User Pipeline \citep{dowell10} using HIPE version 15.01, using the standard SPIRE Calibration Tree 14.3 with default values for all pipeline tasks for processing the detector timelines from the Level 0 raw data to the Level 1 flux calibrated timelines. The timelines were then baseline-subtracted / destriped using the standard SPIRE pipeline destriper tool to remove residual temperature drifts. Final individual  point source maps, calibrated in Jy beam$^{-1}$, for each observation were produced using the standard SPIRE pipeline naive mapper tool  with pixel sizes of 6$\arcsec$, 10$\arcsec$ and 14$\arcsec$ at 250$\, \umu$m, 350$\, \umu$m and 500$\, \umu$m, respectively. Scan-map turnarounds were also included in order to optimise the areal coverage and to improve the map edges.

In order to create a final combined map of all the SPIRE Dark Field observations, the individual maps must be stacked and merged. In general, SPIRE maps are accurate to around 2$\arcsec$, however telescope jitter and other effects may introduce astrometric offsets of up to a pixel. Therefore, in order to merge the Dark Field observations, the astrometry of each individual map must be aligned. The HIPE processing environment provides both astrometry correction and map merging tools at both the Level 1 timeline and the Level 2 image map stages. Both the Level 2 and Level 1 timeline astrometry correction tools were tested. The tools assume a reference SPIRE observation (in this case observation ID  1342192082) aligning the other observations to this reference. The effectiveness of both tools was validated by extracting sources from the final aligned maps and comparing their positions. The astrometry correction using the timeline data was ultimately selected since the source separations were consistently lower than  the image map level method. A median shift of 1.69$\arcsec$ was found, with only 3/141 observations having a shift $>$5$\arcsec$ (still less than the PSW band pixel scale). These three observations were manually checked and confirmed to be valid observations. 

\begin{figure*}
  \includegraphics[width=0.7\textwidth]{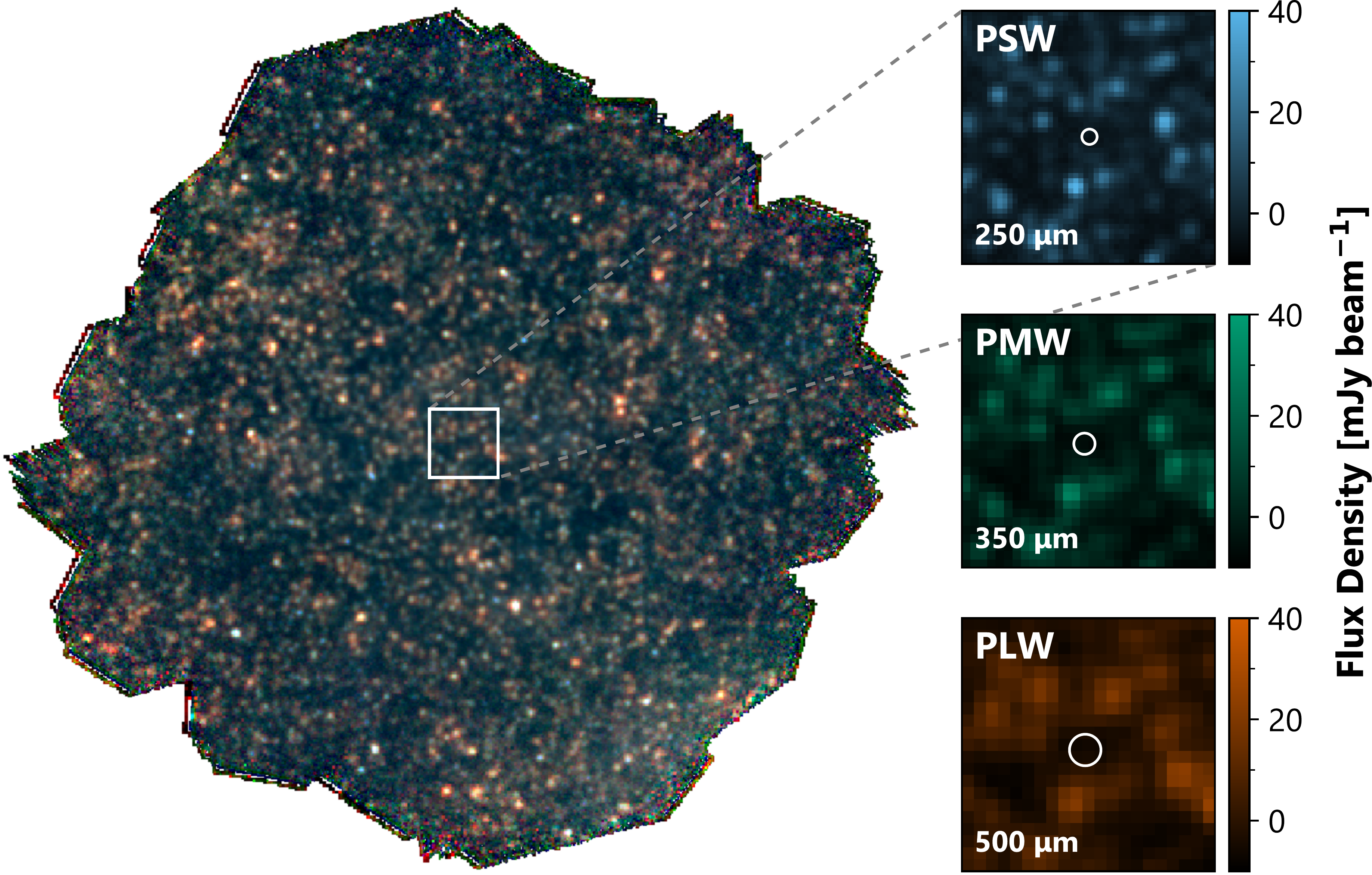}
        \caption{
The 3-colour map of the SPIRE Dark Field after stacking of individual maps in each band and then combining the resulting maps in the SPIRE PSW, PMW, PLW bands. Also shown are cutouts of a central 5$\arcmin$$\times$5$\arcmin$ region in the individual PSW, PMW and PLW bands. The circles on the cutouts correspond to the SPIRE beam size at each wavelength.
        }
  \label{fig:darkfieldRGB}
\end{figure*}

\begin{figure*}
  \includegraphics[width=1.0\textwidth]{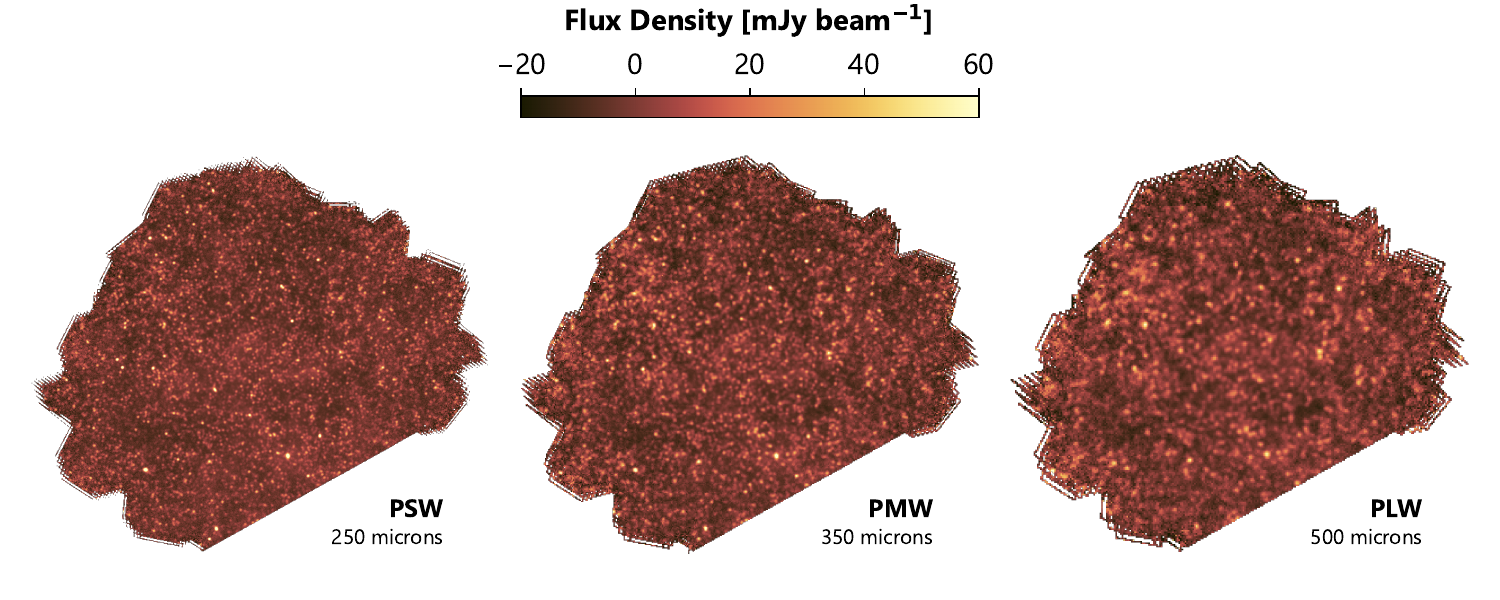}
        \caption{
	Final SPIRE Dark Field point source maps (mJy beam$^{-1}$) in the PSW, PMW \& PLW bands after astrometry alignment, stacking  and masking to remove contamination due to Galactic cirrus.
        }
  \label{fig:darkfieldbands}
\end{figure*}

\begin{figure}
  \includegraphics[width=0.5\textwidth]{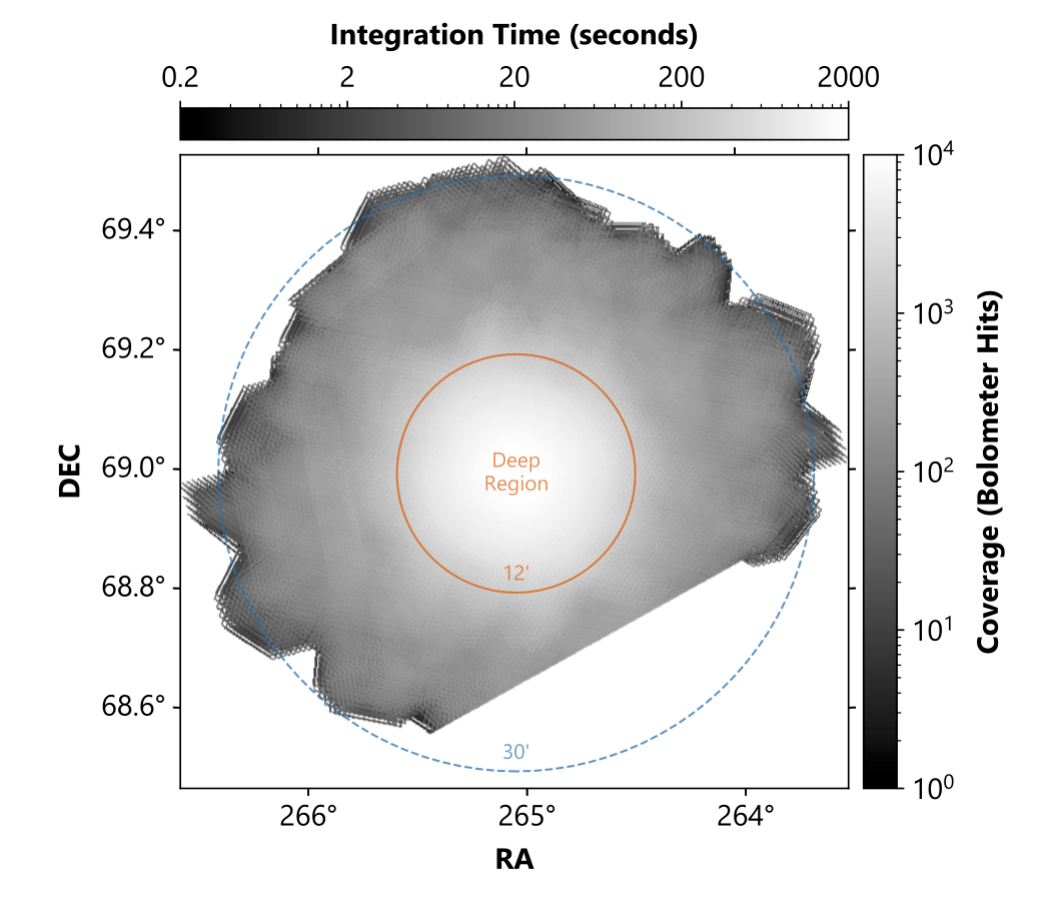}
        \caption{
	Coverage Map of the SPIRE Deep Field in the PSW (250$\, \umu$m) band. The vertical colour bar shows the SPIRE bolometer detector hits per pixel converted into an exposure time on the horizontal colour bar calculated via $pixel size (\arcsec) / scan speed (\arcsec s^{-1}) \times coverage$. Two clear regions can be identified. A deep circular central region of diameter 12$\arcmin$ corresponding to the  SPIRE Small Map mode calibration observations and a shallower surrounding area of diameter $\sim$30$\arcmin$ corresponding to the SPIRE Large Map mode calibration observations.
        }
  \label{fig:coverage}
\end{figure}

The final Dark Field merged map was then assembled using the entire astrometry corrected Level 1 timeline dataset. The entire dataset was baseline corrected using the SPIRE destriper tool and the final map created using the naive mapmaker tool. In Figure~\ref{fig:darkfieldRGB}, the final 3-colour map of the Dark Field is shown approximately $\sim$30$\arcmin$ in diameter, combining the SPIRE PSW, PMW and PLW bands (i.e. 250$\, \umu$m, 350$\, \umu$m and 500$\, \umu$m).  Also shown in  Figure~\ref{fig:darkfieldRGB} are cutouts of a central 5$\arcmin$$\times$5$\arcmin$ region in the individual PSW, PMW and PLW bands. The cutouts clearly show the depth of the field and the crowded nature of the sources. We note broad band diffuse emission in the bottom right hand corner of the map in Figure~\ref{fig:darkfieldRGB} possibly due to contamination by Galactic Cirrus \citep{low84}. Therefore, this region is masked out to produce the final maps in each band shown in Figure~\ref{fig:darkfieldbands}. The coverage map in the PSW (250$\, \umu$m) band is shown in Figure~\ref{fig:coverage} indicating two distinct depths for our Dark Field area. An inner deeper region defined by a circle of radius approximately 6$\arcmin$ corresponding to the stacked SPIRE Small Map mode calibration observations has a coverage of 7000-10,000 SPIRE bolometer detector hits per pixel. An outer region that corresponds to calibration observations taken in  the SPIRE Large Map mode that corresponds to many 100s - 1000s of SPIRE bolometer detector hits per pixel. An integration time per map pixel can be calculated via; $pixel size (\arcsec) / scan speed (\arcsec s^{-1}) \times coverage$ which yields central exposure times of up to 2000 s per pixel. 

\begin{figure}
  \includegraphics[width=0.5\textwidth]{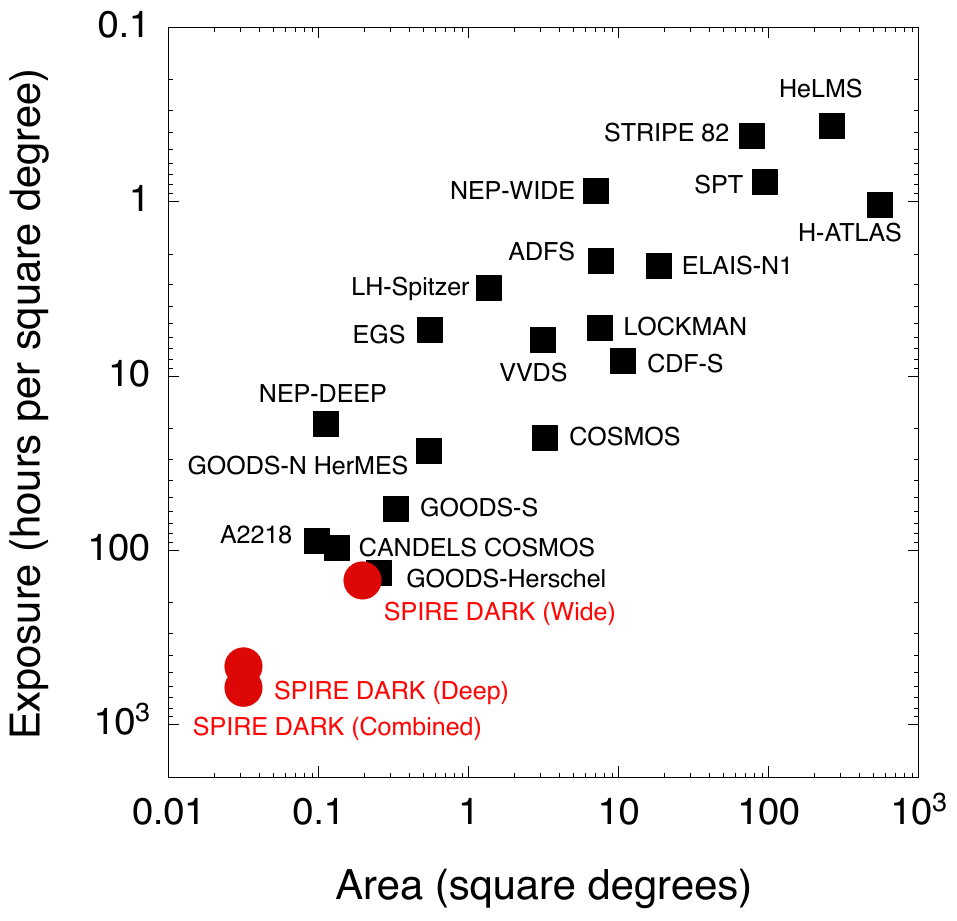}
        \caption{
	 Depth-Area relation for major {\it Herschel} surveys including HerMES \citep{oliver12}, {\it Herschel}-Stripe 82 / HeLMS \citep{viero14}, H-ATLAS \citep{eales10}, GOODS-{\it Herschel} Key Programme \citep{elbaz11}, CANDELS COSMOS \citep{schreiber15}, SPT \citep{holder13}, NEP \citep{pearson17}. Following \citet{lutz14}, the survey area is compared against the exposure time per square degree. The exposure time is calculated from the {\it Herschel} Science Archive and includes all overheads for the observation. The GOODS-Herschel field is the deepest field for extragalactic science to date. The SPIRE Dark Field is decomposed into the wider area, with a similar area and depth to the deepest GOODS field, the smaller, central deeper area as described in Figure~\ref{fig:coverage} and a combined result. The {\it Deep} and {\it Combined} observations produce the deepest {\it Herschel}-SPIRE observations ever undertaken.  
        }
  \label{fig:deptharea}
\end{figure}

In order to compare the depth of the SPIRE Dark Field with other {\it Herschel}-SPIRE surveys we compute an exposure time per survey area (hours / square degree). In Figure~\ref{fig:deptharea}, following \citet{lutz14}, the relative exposure time to survey area is shown against the survey area for a representative sample of {\it Herschel}-SPIRE galaxy field surveys. The exposure time / area metric provides an independent measure of the depth of each survey without taking into account the confusion noise due to background point sources.   The SPIRE surveys cover a wide range of depths and areas from wide-shallow surveys such as H-ATLAS covering 550 square degrees to relatively shallow depths of 30mJy \citep{eales10} to deep-narrow fields such as the {\it Herschel}-GOODS survey, reaching down to the confusion limit. To date, the single deepest extragalactic field survey was the GOODS-{\it Herschel} Key Programme in the GOODS-N field \citep{elbaz11}. This programme consisted of 38 SPIRE scan map observations each of 3 repetitions over a mapping area of 900$\arcmin ^2$  with a total exposure time of just over 30 hours.
As shown in Figure~\ref{fig:coverage}, the SPIRE Dark Field consists of concurrent wide and deep areas.  The wide area, consisting of some larger maps, corresponds to an exposure / area value comparable to the GOODS-Key Programme, the deeper region of the dark field (SPIRE Dark (Deep) in Figure~\ref{fig:deptharea}), created by many repetitions in small map mode reaches an exposure time per square degree significantly deeper than the GOODS-Key Programme. Even combining the GOODS-Key Programme with the shallower HerMES survey in the GOODS-N region, the deepest area of the SPIRE Dark Field  is still 50$\%$ deeper again than the GOODS fields combined, making the SPIRE Dark Field the deepest far-infrared/submillimetre observations ever produced at these wavelengths.

\smallskip

The SPIRE Dark Field consists of a total of 581 scan map repetitions (c.f. 144 repetitions combined in the GOODS-N field) with the noise reducing in proportion to the square root of the number of repetitions.  Given the expected 1$\sigma$ single scan instrument sensitivities of 9.0, 7.5 and 10.8 mJy at 250, 350 and 500$\, \umu$m respectively\footnote{ \tt {http://herschel.esac.esa.int/Docs/SPIRE/spire$\_$handbook.pdf}}, the depth of the SPIRE Dark Field easily integrates down to the measured SPIRE confusion noise due to extragalactic point sources of 5.8, 6.3 and 6.8 mJy at 250, 350 and 500$\, \umu$m respectively as measured by \citet{nguyen10}. The noise in our field was estimated  by splitting the map into equal 2$\arcmin$ $\times$ 2$\arcmin$ regions, distributed across the field. For each region, the standard deviation of the pixel values was calculated to create a pixel flux density distribution. The median and standard deviation of this distribution was then measured to estimate the noise  in the field. This process was repeated for both the entire Dark Field map and the deeper central region in Figure~\ref{fig:coverage}. The resulting noise measurements are shown in Table~\ref{tab:noise} and are consistent with the results of \citet{nguyen10} showing that we are confusion limited in both regions of the Dark Field.

\begin{table}
\caption{Noise estimates for the SPIRE Dark Field over the entire region and the deeper central region compared with the SPIRE confusion noise estimates of \citet{nguyen10}.}
\centering
\begin{tabular}{@{}llll}
\hline
SPIRE Band &  \multicolumn{3}{c}{Estimated Noise (mJy)}    \\
                    &     Full Map    &      Deep Region     &     Confusion	\\
\hline
PSW 250$\, \umu$m               &     5.87 $\pm$ 2.39    &     5.33 $\pm$ 1.75     &    5.8 $\pm$ 0.3	\\ 
PMW 350$\, \umu$m               &    6.05 $\pm$ 2.12  &     5.80 $\pm$ 1.62     &    6.3 $\pm$ 0.4	\\ 
PLW 500$\, \umu$m               &    5.82 $\pm$ 1.88   &   5.34 $\pm$ 1.60    &    6.8 $\pm$ 0.4	\\ 
\hline
\end{tabular}
\label{tab:noise}
\end{table}

\section{Map Based Source Extraction}\label{sec:SourceExtraction}
\begin{figure}
 \includegraphics[width=0.5\textwidth]{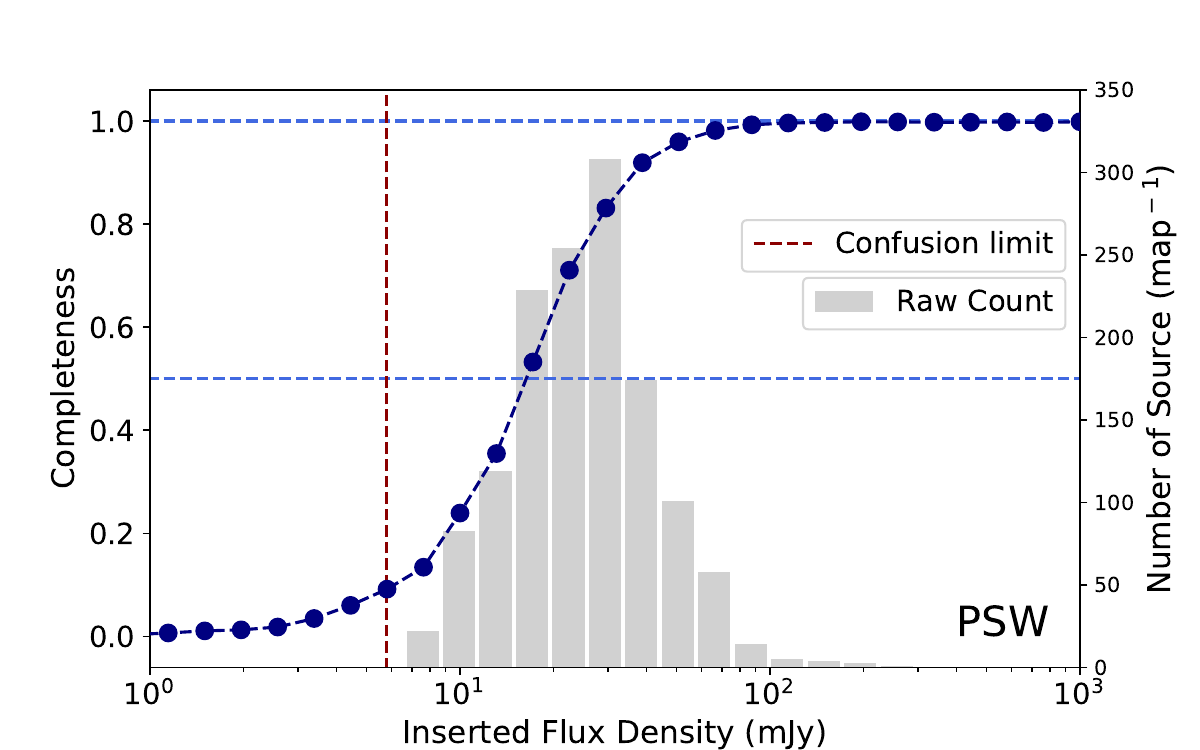}
 \includegraphics[width=0.5\textwidth]{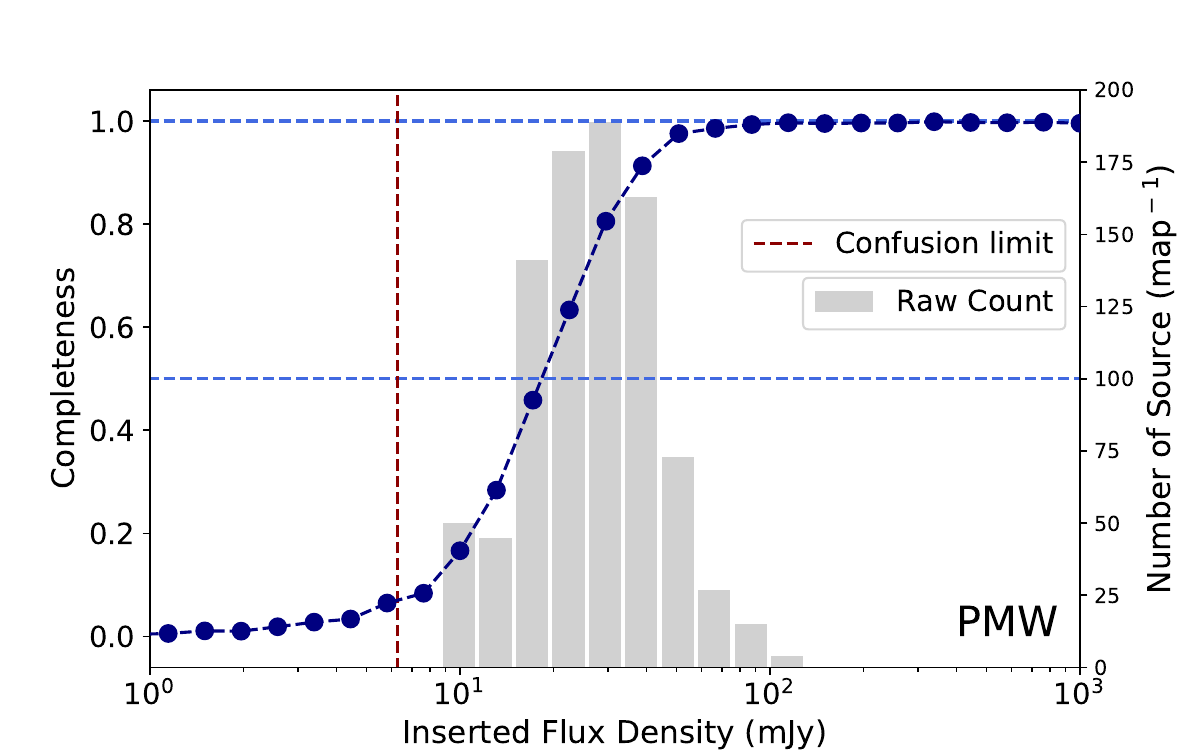}
 \includegraphics[width=0.5\textwidth]{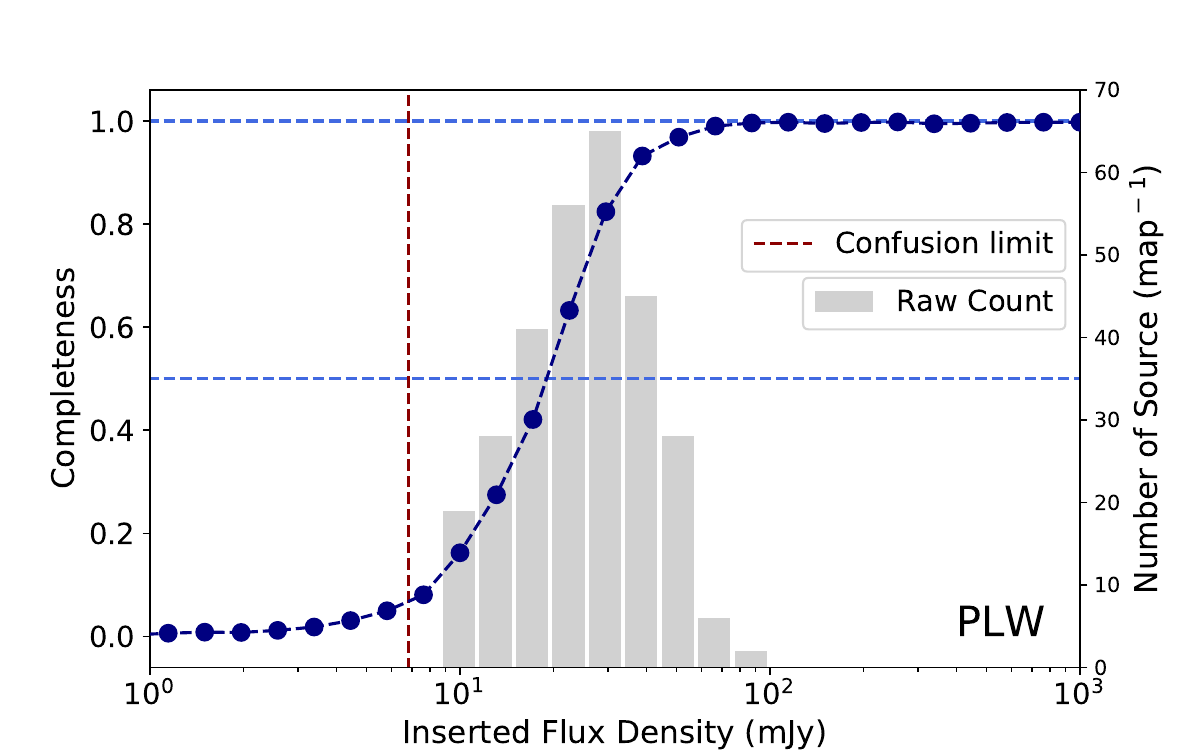}
        \caption{
	 Raw source counts and completeness correction for the PSW (250$\, \umu$m, {\it top-panel}), PMW (350$\, \umu$m {\it middle-panel}) \& PLW (500$\, \umu$m {\it bottom-panel}) bands.  The horizontal dashed lines mark the 50$\%$ and 100$\%$ completeness limits respectively. The vertical dashed line marks the confusion limit in each band. 
        }
  \label{fig:completeness}
\end{figure}

 Initial map based source extraction  was performed using the standard SUSSEXtractor source detection tool available within the HIPE software environment (see Section~\ref{sec:xid} for the results from list driven source extraction).  SUSSEXtractor adopts a Bayesian approach to source detection, modelling both the source and the empty sky at each potential source position to determine whether a source is present at that location and is described in detail in \citet{savage07}, \citet{smith12} and the HIPE implementation in \citet{pearson14}. SUSSEXtractor requires as input the SPIRE band FWHM, detection threshold and an assumed point response function (PRF). We assume mean FWHM of 17.9$\arcsec$, 24.2$\arcsec$ and 35.4$\arcsec$  for the PSW, PMW and PLW bands respectively, a detection threshold of 3$\sigma$ and use the instrumental PRFs obtained from the SPIRE Calibration Tree.
Performing the source extraction independently on the 3 maps, SUSSEXtractor detects a total of 1591, 1073, 381 sources in the  PSW, PMW and PLW bands respectively. The raw source counts are shown in Figure~\ref{fig:completeness}.

A correction for the completeness of sources extracted from the map must be made in order to correct for missed sources, the non-uniform depth of the field and to calculate an effective area over which sources are successfully detected as a function of flux density. The completeness of our source extraction was estimated via Monte Carlo simulations by injecting artificial Gaussian sources of known flux density into the original map. Artificial sources are inserted into the map at random positions ensuring that any artificial source does not lie within one beam FWHM of any real source. SUSSEXtractor is then used to extract sources from this new map and the extracted source catalogue compared with the original artificial source injected positions. If an extracted source lies within 0.5 FWHM of an artificial source then the artificial source is said to be successfully detected. This process is repeated 2000 times, inserting artificial sources into the original  map. The entire process is then repeated for a range of 35 flux values equally spaced logarithmically from 0.1mJy -- 1Jy. The resulting completeness correction as a function of flux density for the 3 SPIRE bands  is shown in Figure~\ref{fig:completeness}. The 50$\%$ completeness levels of 16, 18.5 and 19 mJy are represented as dashed lines in Figure ~\ref{fig:completeness} for the PSW, PMW and PLW bands respectively.

\begin{figure*}
  \includegraphics[width=0.8\textwidth]{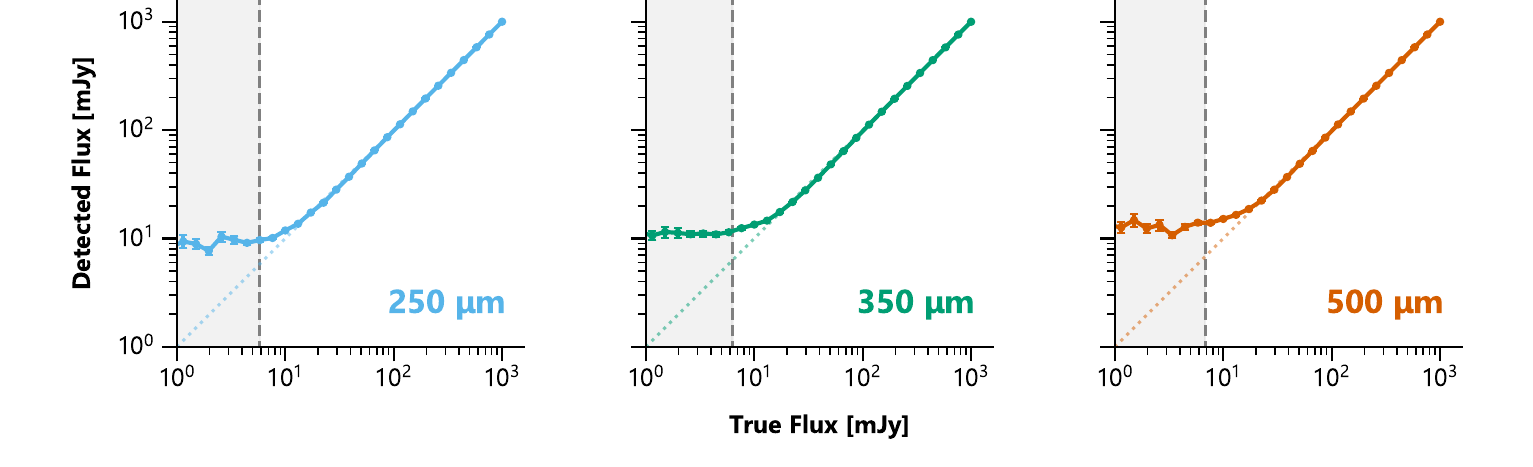}
        \caption{
Comparison of extracted versus true flux densities for simulated sources in the PSW (250$\, \umu$m), PMW (350$\, \umu$m) \& PLW (500$\, \umu$m) bands. Dotted line and shaded region represent the SPIRE confusion noise limit from \citet{nguyen10}. Any deviation from the 1:1 truth line is indicative of flux boosting.
        }
  \label{fig:fluxboosting}
\end{figure*}

Our Monte-Carlo simulations can also be used to address any possible Eddington bias where scatter in the flux density of faint sources can effectively boost the number of brighter sources at flux densities around or above our sensitivity limit. This is particularly pronounced in the longest wavelength 500$\, \umu$m band, due to its large beam size. Using the recovered flux results from the completeness Monte-Carlo simulation, we obtain a correction for any flux boosting by comparing the ratio of recovered versus true source fluxes. The results are shown in Figure~\ref{fig:fluxboosting}. As the 1$\sigma$ confusion noise is approached, we see the relationship between the extracted and true flux values begin to deviate from a linear relationship, slowly flattening out as the confusion limit is reached. Flux boosting corrections are taken into account below flux densities where the relationship deviates from linearity at  $\sim$10, 15, \& 20 mJy in the PSW, PMW and PLW bands respectively. These levels are similar to the 50$\%$ completeness levels.

Finally, in order to quantify the reliability of our source extraction, SUSSEXtractor was also run on the inverted map (i.e. multiplying the image map by -1). A reliability correction on the number of sources as a function of flux density is calculated as (1 - {\it inverted map sources /  nominal map sources}). All SPIRE bands have a $>$80$\%$ reliabilty at flux densities brighter than 20mJy. Below 20mJy the reliability of the PMW ( 350$\, \umu$m) and PLW ( 500$\, \umu$m ) bands drops significantly to approximately 30$\%$ at 10mJy. The PSW (250$\, \umu$m) band remains reliable to fluxes $<$10mJy down to the confusion noise.

\section{Source Counts}\label{sec:SourceCounts}
\begin{figure}
 \includegraphics[width=0.4\textwidth]{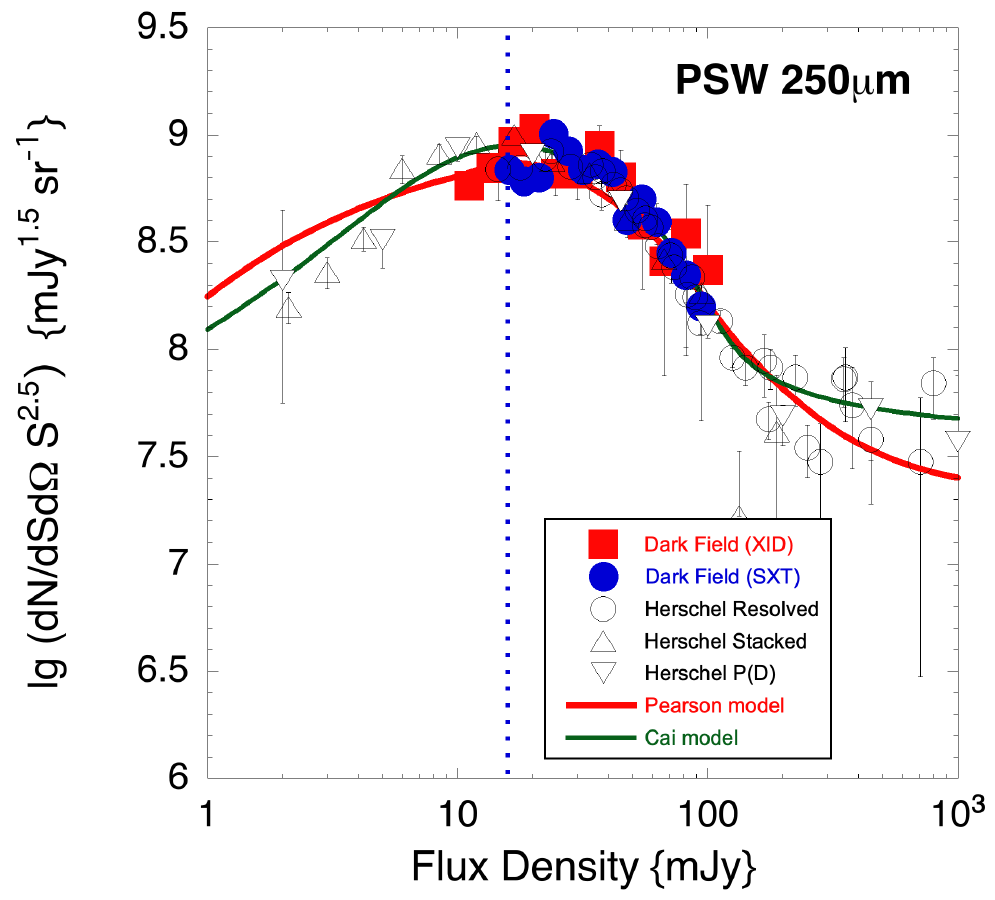}
  \includegraphics[width=0.4\textwidth]{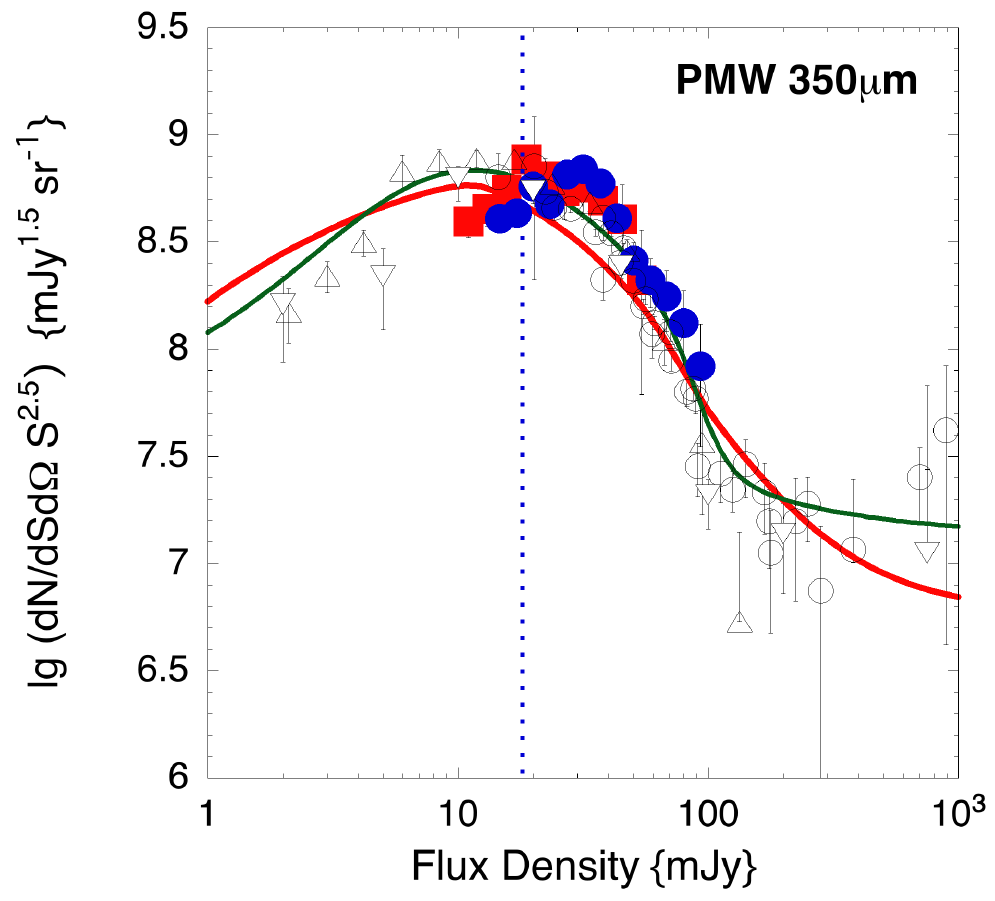}
  \includegraphics[width=0.4\textwidth]{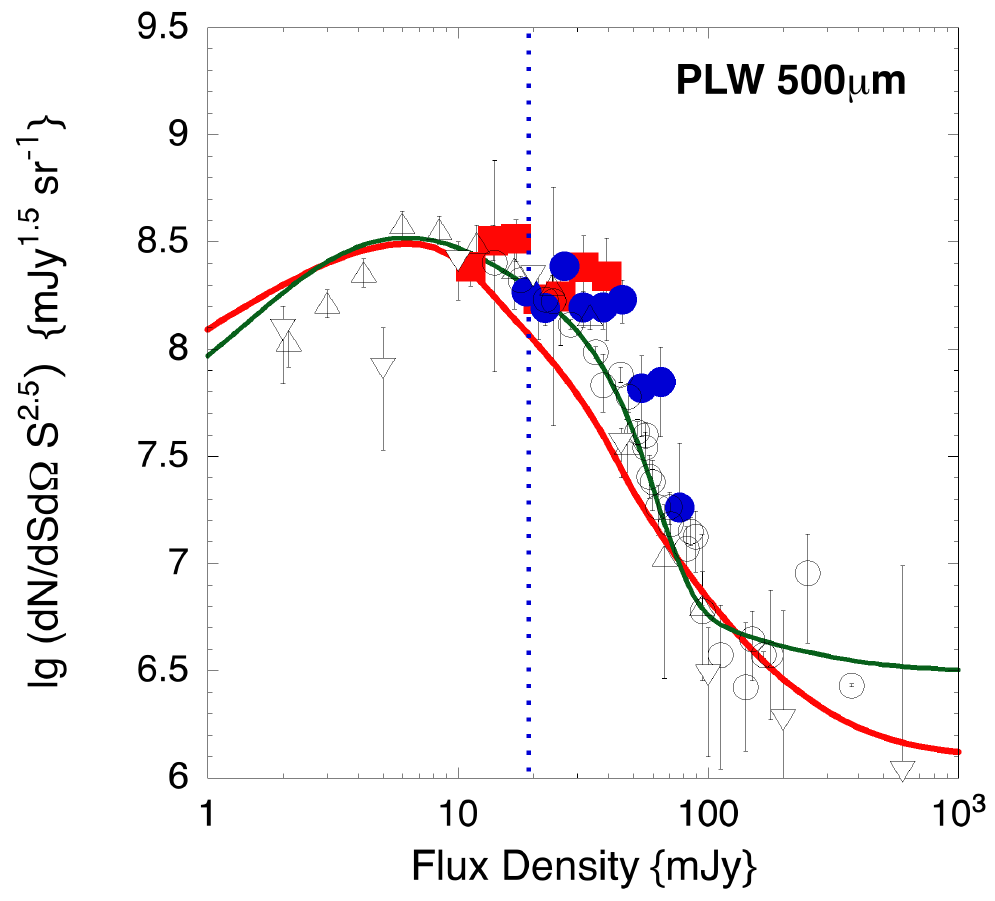}
        \caption{
 SPIRE Dark Field Euclidean normalised differential source counts in the PSW 250$\, \umu$m, PMW 350$\, \umu$m \& PLW 500$\, \umu$m bands derived from the SUSSEXtractor source extraction described in Section~\ref{sec:SourceExtraction} and XID source extraction described in Section~\ref{sec:xid}.  The 50$\%$ completeness limit for the SUSSEXtractor method is shown as a vertical line. Also shown are the resolved source counts from other {\it Herschel} surveys;  HerMES \citep{oliver10}, H-ATLAS  \citep{clements10}, SPIRE-NEP  \citep{pearson17} and the SHARC-II counts at 350$\, \umu$m  \citep{pearson09}, the unresolved stacked counts from \citet{bethermin12} and the P(D) analysis of \citet{glenn10}. Also shown are the galaxy evolution models of  \citet{pearson19} and \citet{cai13}. 
        }
  \label{fig:sourcecounts}
\end{figure}

 In Figure~\ref{fig:sourcecounts} the  Euclidean normalised differential source counts, d$N$/d$S\, \cdot S^{2.5}$ in mJy$^{1.5}$Sr$^{-1}$, derived from our extracted sources are shown for the PSW (250$\, \umu$m), PMW (350$\, \umu$m) \& PLW (500$\, \umu$m) bands.  The integral counts (per square degree) and the differential counts (per steradian) are tabulated in Table~\ref{tab:sourcecounts} assuming logarithmic flux bins of uniform width. Note that the counts in Figure~\ref{fig:sourcecounts}  have been cut at the 50$\%$ completeness limit in each SPIRE band. The source counts have been corrected for the effects of completeness, reliability and flux-boosting. Also shown are results from the major large area surveys carried out with  {\it Herschel}, the HerMES {\it Herschel} Guaranteed Time programme  \citep{oliver10}, H-ATLAS {\it Herschel}  Open Time Key Programme  (\citealt{clements10}, \citealt{valiante16}) and the Open Time observations at the North Ecliptic Pole (NEP, \citealt{pearson17}), plus the SHARC-II 350$\, \umu$m counts from \citet{pearson09}. 
 
 The source counts cover the range from 100mJy - 10mJy in each SPIRE band and are in broad agreement with the results from other fields exhibiting a steep rise to fainter fluxes. The counts are clearly limited by both the small field size (effectively the size of the SPIRE small scan map with a diameter of 12$\arcmin$) and the source confusion due to the depth of the maps. This is particularly evident at 500$\, \umu$m where the counts are rather non-uniform with a large scatter even after correcting for completeness and flux-boosting. In the next section we pursue an alternative methodology for  extracting sources and constructing the source counts in deep confused fields.

\begin{table}
\caption{Source counts derived from the SUSSEXtractor source extraction described in Section~\ref{sec:SourceExtraction}, for the SPIRE PSW, PMW, PLW bands, as plotted in Figure ~\ref{fig:sourcecounts}. Integral counts are number per square degree. Differential source counts normalised to a Euclidean universe, d$N$/d$S\, S^{2.5}$ in mJy$^{1.5}$sr$^{-1}$,  Columns are the flux density (uniform logarithmic bins), counts and errors.
}
\centering
\begin{tabular}{@{}lllll}
\hline
Flux  &  Integral  &  Errors &  Differential  &  Errors   \\
mJy   &  \multicolumn{2}{c}{deg.$^{-2}$}&  \multicolumn{2}{c}{mJy$^{1.5}$sr$^{-1}$}   \\
  &  & & \multicolumn{2}{c}{$\times$10$^7$}    \\
\hline
 \multicolumn{5}{c}{PSW 250 $\, \umu$m Band}     \\
  \hline
10.7 & 3098.4 & 55.7 & 26.677 & 3.533 \\
12.3 & 2785.0 & 52.8 & 32.125 & 3.867 \\
14.1 & 2477.0 & 49.8 & 39.141 & 4.271 \\
16.1 & 2170.8 & 46.6 & 68.561 & 5.694 \\
18.4 & 1733.0 & 41.6 & 60.637 & 5.402 \\
21.1 & 1416.9 & 37.6 & 62.982 & 5.679 \\
24.2 & 1149.1 & 33.9 & 100.851 & 7.559 \\
27.7 & 799.0 & 28.3 & 83.822 & 7.324 \\
31.7 & 561.5 & 23.7 & 68.737 & 7.128 \\
36.3 & 402.6 & 20.1 & 73.134 & 7.932 \\
41.6 & 264.5 & 16.3 & 66.986 & 8.245 \\
47.6 & 161.4 & 12.7 & 40.223 & 7.002 \\
54.5 & 110.8 & 10.5 & 50.050 & 8.584 \\
62.4 & 59.5 & 7.7 & 39.262 & 8.371 \\
71.4 & 26.6 & 5.2 & 28.183 & 7.817 \\
81.8 & 6.9 & 2.6 & 22.131& 7.911 \\
93.7 & 0.28& 1.1 & 15.849 & 7.827 \\
\hline
 \multicolumn{5}{c}{PMW 350 $\, \umu$m Band}     \\
 \hline
10.8 & 2123.5 & 46.1 & 12.366 & 2.579 \\
12.6 & 1961.2 & 44.3 & 17.140 & 3.030 \\
14.7 & 1782.6 & 42.2 & 40.434 & 4.465 \\
17.2 & 1447.8 & 38.1 & 43.016 & 4.560 \\
20.0 & 1164.9 & 34.1 & 57.264 & 5.317 \\
23.3 & 865.8 & 29.4 & 47.266 & 4.982 \\
27.2 & 669.7 & 25.9 & 65.235 & 6.164 \\
31.7 & 454.7 & 21.3 & 68.823 & 6.716 \\
37.0 & 274.5 & 16.6 & 59.433 & 6.773 \\
43.1 & 150.9 & 12.3 & 40.791 & 6.149 \\
50.3 & 83.6 & 9.1 & 25.925 & 5.406 \\
58.6 & 49.5 & 7.0 & 20.996 & 5.421 \\
68.3 & 27.7 & 5.3 & 17.626 & 5.574 \\
79.7 & 13.1 & 3.6 & 13.276 & 5.420 \\
92.9 & 4.3 & 2.1 & 8.317 & 4.802 \\
 \hline
 \multicolumn{5}{c}{PLW 500 $\, \umu$m Band}     \\
 \hline
11.0 & 766.9 & 27.7 & 7.370 & 1.903 \\
13.1 & 657.5 & 25.6 & 10.127 & 2.159 \\
15.6 & 542.2 & 23.3 & 12.766 & 2.413 \\
18.7 & 430.7 & 20.8 & 18.502 & 2.890 \\
22.3 & 306.9 & 17.5 & 15.563 & 2.669 \\
26.6 & 227.0 & 15.1 & 24.322 & 3.440 \\
31.7 & 131.3 & 11.5 & 15.629 & 2.954 \\
37.9 & 84.2 & 9.2 & 15.647 & 3.263 \\
45.2 & 48.0 & 6.9 & 17.058 & 3.814 \\
54.0 & 17.8 & 4.2 & 6.599 & 2.694 \\
64.5 & 8.8 & 3.0 & 7.051 & 3.153 \\
77.0 & 1.5 & 1.2 & 1.819 & 1.819 \\
\hline
\end{tabular}
\label{tab:sourcecounts}
\end{table}

\section{List driven source extraction from 24$\, \umu$\lowercase{m} XID}\label{sec:xid}

The depth of the SPIRE Dark Field makes source extraction challenging due to the maps reaching below the confusion limit and the large associated SPIRE beams which can produce blending between sources. An alternative to our methodology using SUSSEXtractor to extract sources from the image map, is list driven source extraction, using a deep ancillary catalogue at a different wavelength for the source position and then performing photometry at the ancillary source position (\citealt{scott02}, \citealt{magnelli09}, \citealt{roseboom10}, \citealt{chapin11}). 

For this method we use the {\it Simultaneous Source Extractor} XID tool within the HIPE environment. XID uses a matrix inversion technique  to fit for the flux density of all known sources simultaneously, assuming priors based on an ancillary input list (\citealt{roseboom10}, \citealt{hurley16}). The observational data set (pixel flux, noise) is encoded in a linear equation and a maximum likelihood method is used to fit a model of the source fluxes to the observations, given positional priors from the ancillary catalogue. For example, for a SPIRE source, with two candidates at the ancillary data wavelength, XID varies the contribution of the fluxes for each of these sources until they accurately replicate the observed SPIRE data. To efficiently calculate this optimal solution, a matrix-inversion technique is employed to solve the maximum likelihood linear equation. Then MCMC methods can be used to efficiently map the posterior distribution.

\begin{figure}
  \includegraphics[width=0.5\textwidth]{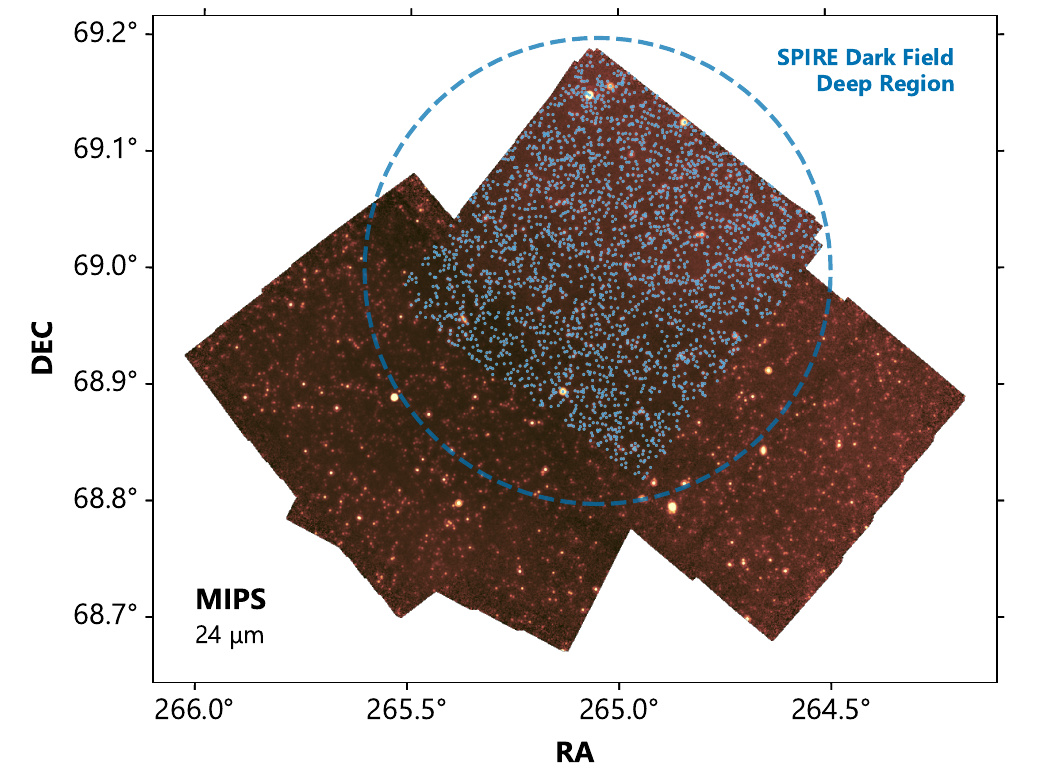}
        \caption{
	 The {\it Spitzer}-MIPS image at  24$\, \umu$m showing the coverage of the SPIRE Dark Field. The white circles correspond  to the sources in the MIPS 24$\, \umu$m catalogue.
        }
  \label{fig:mips}
\end{figure}

Due to the strong correlation with the sub-millimetre emission (\citealt{elbaz10}, \citealt{bethermin12}), mid-infrared data is best suited to the XID process. Therefore, the {\it Spitzer} MIPS  24$\, \umu$m data covering the IRAC Dark Field \citep{krick09} was selected on the basis of its superior depth and resolution. The NASA/IPAC Infared Science Archive (IRSA\footnote{ \tt {https://irsa.ipac.caltech.edu/data/SPITZER/} \\ \tt {docs/spitzerdataarchives/}}) was searched for MIPS 24$\, \umu$m images at the position of the SPIRE Dark Field and the 24$\, \umu$m (multi-wavelength catalogue) was obtained from \citet{krick09}. In Figure~\ref{fig:mips}, the MIPS images and 24$\, \umu$m catalogue are shown with the SPIRE Dark Field deep region (12$\arcmin$ in diameter) overplotted. Although the MIPS image and catalogue only partially cover the SPIRE Dark Field they still provide an excellent dataset with which to perform our XID analysis.

\begin{figure}
  \includegraphics[width=0.5\textwidth]{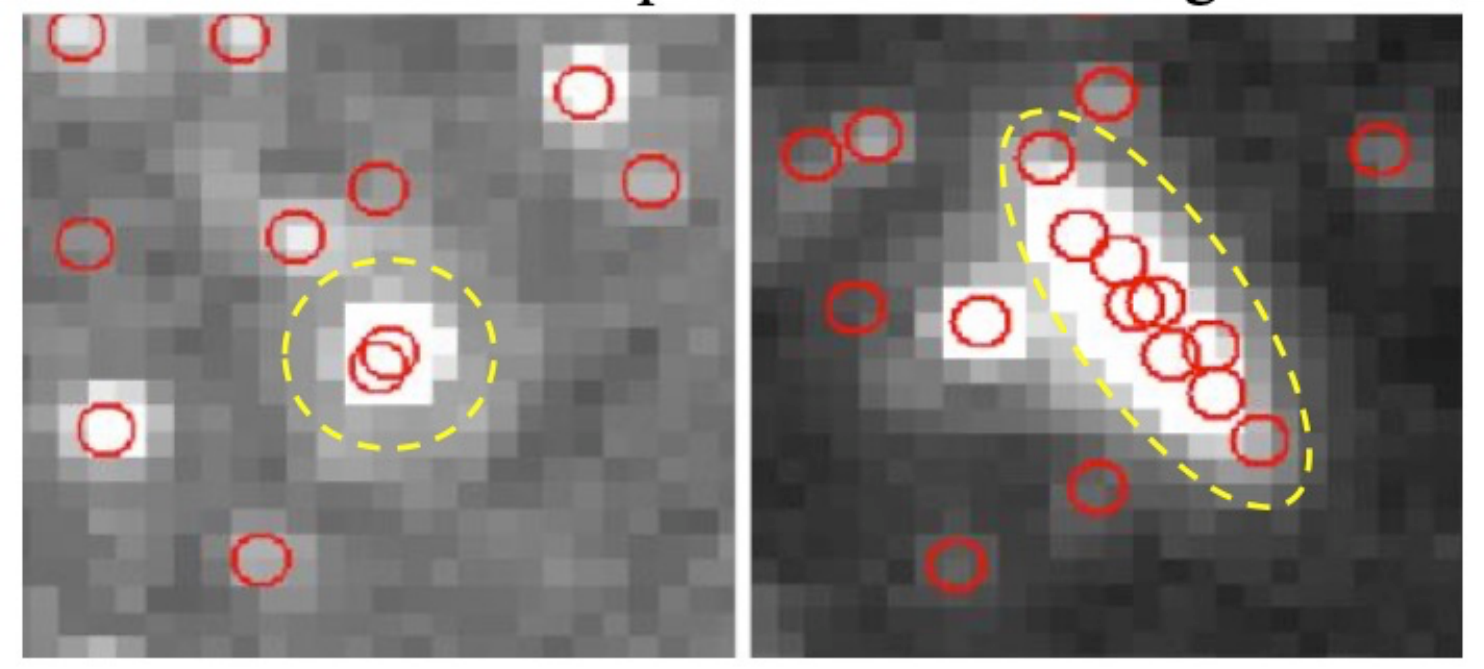}
        \caption{
	The {\it Spitzer}-MIPS catalogue overlaid on the MIPS-24$\, \umu$m image, {\it left}: multiple detections of the same source, {\it right}: shredding of an extended source into multiple MIPS sources. These types of anomalies are reduced by flux cuts, analysis of signal-to-noise and visual inspection of the MIPS images in order to produce a robust final catalogue at 24$\, \umu$m.
        }
  \label{fig:mipsproblems}
\end{figure}

The initial  MIPS  24$\, \umu$m catalogue contains approximately 3300 entries covering an area of  225$\arcmin^{2}$ (3$\times$3 MIPS fields of view). However, on closer inspection, 20$\%$ of the detections have a signal-to-noise (S/N) $<$ 5$\sigma$. Moreover, on overlaying the MIPS catalogue on the MIPS image we find many multiple detections and also cases where an extended source has been shredded into multiple point sources as shown in the left and right panels of Figure~\ref{fig:mipsproblems}. Therefore, in order to mitigate these effects and ensure a robust input catalogue for our  {\it Simultaneous Source Extractor} XID analysis, the original 24$\, \umu$m catalogue is filtered on the following criteria; we discard any sources with negative fluxes or any with S/N $<$ 5$\sigma$. For each source, we then identify any additional source within 1 MIPS 24$\, \umu$m FWHM (5.9$\arcsec$), assuming the brightest detection as the source. Finally, we again iterate through the filtered catalogue, matching sources within 0.5 FWHM of one another. Double-detections are identified, and the brightest of the pair is kept. The final filtered catalogue contains 2625 robust MIPS 24$\, \umu$m detections.

This final MIPS catalogue provides the input to the {\it Simultaneous Source Extractor} XID process. The MIPS catalogue was sorted from brightest to faintest flux density. The {\it Simultaneous Source Extractor} XID process then iterated through this catalogue in decreasing brightness. At each iteration, the extracted sources were subtracted from the SPIRE map using the SPIRE PRF from the SPIRE Calibration Tree in the appropriate band. The process is then repeated until the input MIPS catalogue has been exhausted. The final SPIRE XID catalogue contains 1848, 1609, 1208  sources with positive flux density in the SPIRE PSW 250$\, \umu$m, PMW 350$\, \umu$m and PLW 500$\, \umu$m bands respectively. These sources all have 24$\, \umu$m counterparts and ancillary data at other wavelengths (e.g. {\it Spitzer} IRAC).

\begin{figure}
 \includegraphics[width=0.5\textwidth]{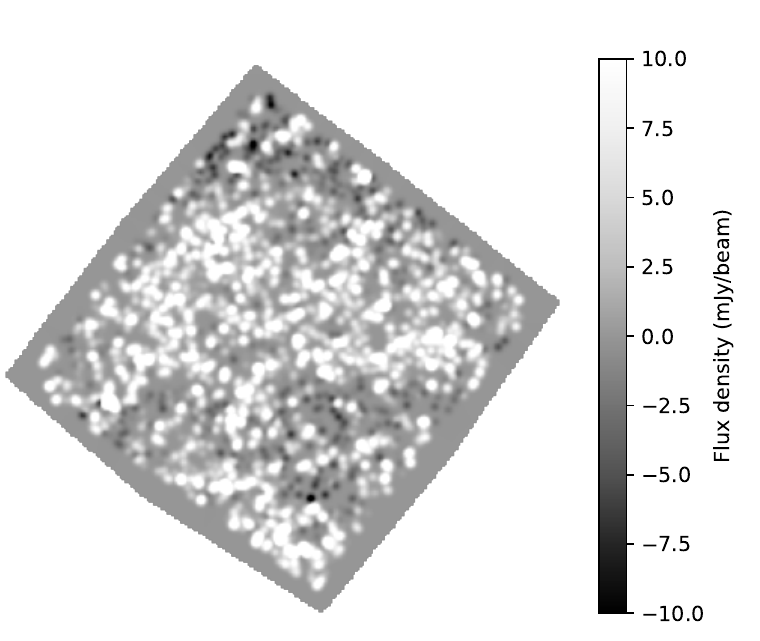}
\includegraphics[width=0.5\textwidth]{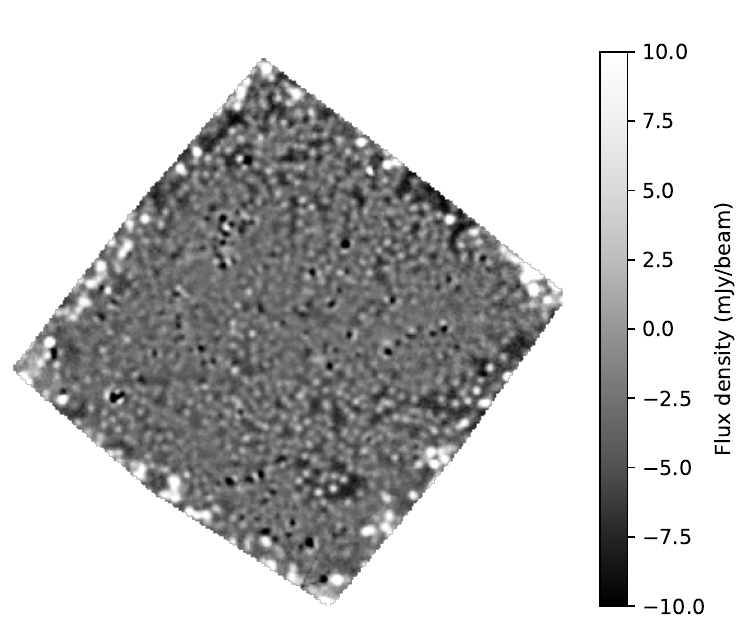}
        \caption{
	 {\it Top}: Source detection map from the {\it Simultaneous Source Extractor} XID tool for the PSW map containing only extracted sources.  {\it Bottom}: The PSW residual map, i.e. the PSW map with MIPS-identified sources removed. Colour bars show the signal level in mJy. The residual map appears approximately uniform within the region covered by MIPS, with a background signal well below the confusion limit. 
	         }
  \label{fig:mipsresiduals}
  \end{figure}

Figure~\ref{fig:mipsresiduals} shows the final source detection map in the SPIRE PSW band, containing only extracted sources, and the residual map, after extraction of sources. The residual map appears approximately uniform within the region covered by MIPS. The source counts derived from the XID process are shown in Table~\ref{tab:xidcounts} for both the integral counts (per square degree) and the differential counts (per steradian) and plotted in Figure~\ref{fig:sourcecounts}.

\begin{table}
\caption{
Source counts derived from the XID simultaneous source extraction described in Section~\ref{sec:xid}, for the SPIRE PSW, PMW, PLW bands, as plotted in Figure ~\ref{fig:sourcecounts}. Integral counts are number per square degree. Differential source counts normalised to a Euclidean universe, d$N$/d$S\, S^{2.5}$ in mJy$^{1.5}$sr$^{-1}$,  Columns are the flux density (uniform logarithmic bins), counts and errors. 
}
\centering
\begin{tabular}{@{}lllll}
\hline
Flux  &  Integral  &  Errors &  Differential  &  Errors   \\
mJy   &  \multicolumn{2}{c}{deg.$^{-2}$}&  \multicolumn{2}{c}{mJy$^{1.5}$sr$^{-1}$}   \\
  &  & & \multicolumn{2}{c}{$\times$10$^7$}    \\
\hline
 \multicolumn{5}{c}{PSW 250 $\, \umu$m Band}     \\
  \hline
11.1 & 4435.1 & 66.6 & 58.123 & 7.101 \\
13.6 & 3479.6 & 59.0 & 70.231 & 9.067 \\
16.6 & 2624.0 & 51.2 & 93.182 & 12.131 \\
20.2 & 1782.6 & 42.2 & 106.549 & 15.068 \\
24.7 & 1069.6 & 32.7 & 66.132 & 13.789 \\
30.1 & 741.6 & 27.2 & 65.953 & 15.996 \\
36.8 & 499.1 & 22.3 & 88.990 & 21.583 \\
44.9 & 256.7 & 16.0 & 63.568 & 21.189 \\
54.9 & 128.3 & 11.3 & 38.120 & 19.060 \\
67.0 & 71.3 & 8.4 & 25.718 & 18.185 \\
81.8 & 42.8 & 6.5 & 34.700 & 24.537 \\
99.9 & 14.3 & 3.8 & 23.410 & 23.410 \\
\hline
 \multicolumn{5}{c}{PMW 350 $\, \umu$mBand}     \\
 \hline
11.0 & 3025.9 & 55.0 & 39.402 & 6.154 \\
13.1 & 2440.7 & 49.4 & 45.126 & 7.521 \\
15.6 & 1926.9 & 43.9 & 55.588 & 9.533 \\
18.7 & 1441.6 & 38.0 & 76.770 & 12.795 \\
22.3 & 927.8 & 30.5 & 63.974 & 13.339 \\
26.6 & 599.5 & 24.5 & 54.419 & 14.051 \\
31.7 & 385.4 & 19.6 & 56.784 & 16.392 \\
37.9 & 214.1 & 14.6 & 49.376 & 17.457 \\
45.2 & 99.9 & 10.0 & 40.252 & 18.001 \\
54.0 & 28.5 & 5.3 & 21.000 & 14.850 \\
  \hline
 \multicolumn{5}{c}{PLW 500 $\, \umu$mBand}     \\
  \hline
11.2 & 1454.2 & 38.1 & 24.187 & 4.491 \\
13.8 & 1040.7 & 32.3 & 31.967 & 6.041 \\
17.0 & 641.6 & 25.3 & 32.819 & 7.162 \\
20.9 & 342.2 & 18.5 & 17.115 & 6.051 \\
25.8 & 228.1 & 15.1 & 17.571 & 7.173 \\
31.8 & 142.6 & 11.9 & 24.052 & 9.819 \\
39.2 & 57.0 & 7.6 & 21.950 & 10.975 \\
  \hline
\end{tabular}
\label{tab:xidcounts}
\end{table}

\section{Results}\label{sec:Results}
The final source counts using the traditional source extraction method (SUSSEXtractor) and the list driven method (XID) are compared in Figure~\ref{fig:sourcecounts}. In general, the XID counts probe approximately a factor of two deeper than the SUSSEXtractor counts and match the depth of the faintest resolved source counts in other {\it Herschel} surveys. The  XID counts reach deep enough to detect the  turnover in the source counts in the 10 - 30mJy range but cannot provide information on what happens after the turnover. In addition to the Dark Field (SUSSEXtractor and XID) and other {\it Herschel} surveys plotted in Figure~\ref{fig:sourcecounts}, we show the results from the stacking analysis in the GOODS-N and COSMOS fields from \citet{bethermin12} and the results of the P(D) analysis from \citet{glenn10}. These counts agree with the results in the Dark Field over the concurrent flux range and extend the restrictions on the source counts to below the SPIRE confusion limit. 

A comparison of the SUSSEXtractor and XID source extraction methods allows an opportunity to  analyse how well the MIPS-24$\, \umu$m sources (used as priors for the XID process) correspond to the light in the SPIRE map. Note, as shown in the coverage map in Figure~\ref{fig:coverage} and MIPS coverage in Figure~\ref{fig:mips}, that the SPIRE field covers a wider area than the corresponding MIPS coverage and we therefore expect more sources at the brighter end of the source counts to be found by  SUSSEXtractor. This is clear in the 350 and 500$\, \umu$m bands in Figure~\ref{fig:sourcecounts} and although it appears that XID finds a similar number of sources at the brightest fluxes in the 250$\, \umu$m band, the errors on the XID counts at these bright fluxes are consistent with zero sources (see Table~\ref{tab:xidcounts}). 

We can assess the extent to which the XID and blind-extracted (SUSSEXtractor) differential counts match over the coincident map area and flux ranges where they overlap, by a flux-flux correlation.  We cross-correlated our SUSSEXtractor and XID catalogues, using a separation corresponding to the beam size in each of the SPIRE bands, to create a catalogue of matched sources with the two corresponding flux measurements. We find that across the range of flux densities where the SUSSEXtractor and XID source counts generally overlap ($\sim$20 - 80mJy), the measured fluxes are well correlated with a linear relationship and Pearson correlation coefficient, $r \sim 0.9$ in all SPIRE bands. Across the full range of flux densities, the correlation is weaker with, $r \sim 0.8 - 0.6$  from the SPIRE PSW 250$\, \umu$m to PLW 500$\, \umu$m band respectively.

\begin{figure}
\centerline{
 \includegraphics[width=0.42\textwidth]{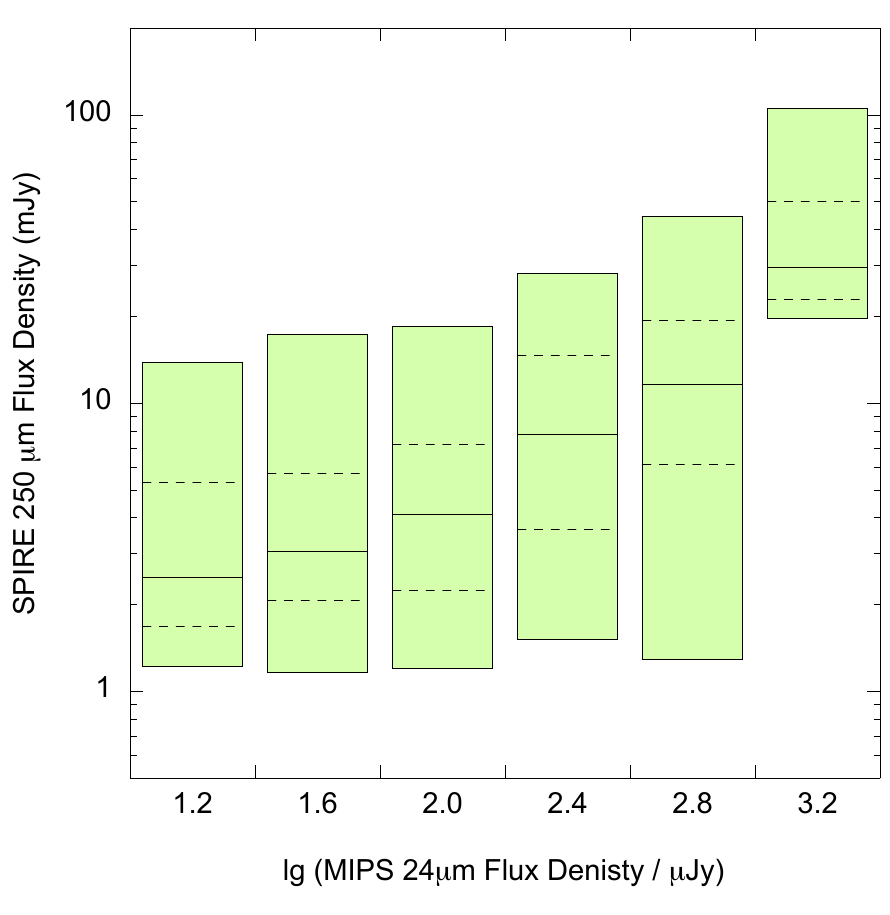}
}
        \caption{
	 Percentile plot of extracted SPIRE fluxes in the PSW  250 $\, \umu$m band as a function of the logarithmic binned input MIPS 24$\, \umu$m prior flux. MIPS 24$\, \umu$m  fluxes on the ordinate and quartile ranges for the SPIRE fluxes on the abscissa. The box limits represent 5$\%$ and 95$\%$ of the data respectively, the solid horizontal lines in the boxes are median, with the upper and lower quartiles shown as dotted lines.
	  	         }
  \label{fig:percentile}
  \end{figure}

For the XID catalogue alone, we also analyse the  distribution of extracted SPIRE fluxes at 250 $\, \umu$m as a function of MIPS 24$\, \umu$m prior flux. The resulting distribution is shown in Figure~\ref{fig:percentile}. The input MIPS fluxes are logarithmically binned and the SPIRE flux distribution is defined by the 5$\%$ and 95$\%$ sample limits, with the median SPIRE flux and upper and lower quartiles shown as solid and dashed horizontal lines respectively.  At the fainter MIPS flux levels, the MIPS sources are yielding SPIRE fluxes around the  confusion limit ($\sim$5.8 mJy), with the median SPIRE flux  increasing to brighter MIPS 24$\, \umu$m fluxes $>$100$\, \umu$Jy.

In order to interpret our results, the predictions of two contemporary evolutionary models, referred to as the Cai and Pearson models are also shown in Figure~\ref{fig:sourcecounts}. The Cai model \citep{cai13} assumes a late (spirals) and early (spheroids) galaxy population but also includes a third population of lensed spheroids from the models of \citet{negrello17}.  The Pearson model is based on the models of \citet{pearson19}. The model incorporates multispectral components from the spectral energy distribution libraries of  \citet{efstathiou00, efstathiou03} with a normal quiescent galaxy component, AGN component and star-forming populations as a function of luminosity from starburst ($\rm L_{IR}\rm <10^{11}L_\odot$), luminous (LIRG;  $\rm 10^{11}L_\odot < L_{IR}< 10^{12}L_\odot$) and ultra-luminous (ULIRG;  $\rm L_{IR}\rm >10^{12}L_\odot$) components. All star-forming components are assumed to evolve in luminosity and number density following the burst scenario introduced in  \citet{pearson05}, with an exponential function to $z \sim$1 and a smooth decline to higher redshift. The AGN population makes a negligible  contribution to the source counts at far-infrared wavelengths.  

Both evolutionary models fit the source counts in the Dark Field at 250$\, \umu$m and 350$\, \umu$m well. At 500$\, \umu$m the Cai model provides a better fit to the steep rise in the counts most probably due to the inclusion of the lensed component.
At the fainter fluxes, the counts are relatively poorly constrained by the stacking \citep{bethermin12} and P(D) results \citep{glenn10} with both models being broadly consistent with the observations. Both models predict the dominant contribution at the fainter fluxes ($<$10 mJy) is made by star-forming galaxies, with the Pearson model specifically predicting a lower luminosity star-forming population.

\smallskip

\section{Conclusions}\label{sec:Conclusions}

We have produced one of the deepest images of the sub-millimetre sky using 141 observations of the SPIRE Dark Field taken during the routine instrument calibration phase of the {\it Herschel} mission. The images consist of all observations in SPIRE large map and SPIRE small map modes taken using the nominal configuration. The final map consists of a roughly 30$\arcmin$ diameter outer region with a deeper central region of  diameter 12$\arcmin$. Both regions reach below the source confusion limits in each of the SPIRE PSW 250$\, \umu$m, PMW 350$\, \umu$m and PLW 500$\, \umu$m bands. 

Sources were initially extracted from the final maps using the SUSSEXtractor algorithm, detecting almost 1800 sources with at least a single detection in any SPIRE band. Of these, 169 sources were without a detection in the shortest wavelength PSW 250$\, \umu$m band, including a number ($\sim$6) of potential high-redshift candidates identified as 500$\, \umu$m risers \citep{dowell14} with detections in more than one SPIRE band. Such high redshift candidates will be the subject of future work.  

The depth of the SPIRE Dark Field means that the maps are highly confused making traditional source extraction challenging, therefore sources were also extracted using a list driven process (XID). Using the {\it Spitzer} MIPS-24$\, \umu$m data in the same field (albeit over a smaller area) as an input catalogue, sources in the SPIRE maps were extracted using a maximum likelihood method. This XID process resulted in 1848 sources with detections in at least one SPIRE band and a corresponding 24$\, \umu$m flux density.

The final source counts were corrected for completeness, reliability and flux-boosting effects and were found to be in good agreement with other {\it Herschel} surveys. The XID process allows us to push the source counts to a factor of two deeper compared to the SUSSEXtractor method. A comparison with two contemporary galaxy evolution models was made showing good agreement with the models over the relevant flux range. 
The Dark Field source counts detect the turnover in the galaxy number distribution but do not extend to flux levels past this peak and therefore do not constrain the counts or models to fainter fluxes. Alternative methods such as stacking or P(D) analysis (\citealt{bethermin12}, \citealt{glenn10}) are required to probe  to these levels and the depth of the Dark Field lends itself well to these methods. The results of  P(D) analysis in the SPIRE Dark Field will be presented in a forthcoming work (Varnish et al. in preparation).

The SPIRE Dark Field provides an important deep dataset with which to investigate galaxy populations and evolution. Its colocation with the {\it Spitzer} IRAC Dark Field means that a wealth of ancillary data from  infrared to X-ray wavelengths already exists, making this a valuable field for future study. The field has already been partially observed with the {\it JWST} \citep{gardner06} at 0.9-4.5$\, \umu$m with the NIRcam instrument as part of the PEARLS programme \citep{windhorst23} and future observations with MIRI would be valuable.

 Future opportunities to observe the SPIRE Dark Field at far-infrared wavelengths may include the potential NASA FIR Probe concepts \citep{vandertak23}, responding to the NASA Astro 2020 Decadal Survey\footnote{ \tt {www.nationalacademies.org/our-work/ \\ decadal-survey-on-astronomy-and-astrophysics-2020-astro2020}}, since no far-infrared concepts are currently envisioned in the ESA Voyage 2050 programme\footnote{ \tt {www.cosmos.esa.int/web/voyage-2050}}. 

The initial down selection for the NASA Probe programme has been concluded with the {\it Probe Far-Infrared Mission for Astrophysics} ({\it PRIMA}, \citealt{bradford23}) proceeding to the Phase-A study.  {\it PRIMA} is a 2 m class aperture and will therefore not provide higher angular resolution than {\it Herschel} at far-infrared red wavelengths. However, the {\it PRIMA} instrumentation includes the FIRESS multimode survey spectrometer \citep{bradford23} with a two order of magnitude increase in mapping speed compared to {\it Herschel} and the PRIMAger Hyper-spectral Imager \citep{ciesla24}. These instruments will provide 40$\, \umu$m to $\sim$250$\, \umu$m spectral data cubes that would allow faint sources to be probed below the confusion limit \citep{clements07}. PRIMAger will also extend to shorter wavelengths of 40-25$\, \umu$m providing the potential for higher resolution priors for longer wavelength images.

Additional Probe proposals not down selected were the {\it Single Aperture Large Telescope for Universe Studies} ({\it SALTUS}, \citealt{kim22}), a 20m inflatable warm (45K) telescope with a suite of spectroscopic instrumentation. SALTUS would have angular resolution of about 1 arcsecond, allowing detailed followup of individual sources but limited in respect to field studies. The {\it Far-Infrared Spectroscopy Space Telescope} ({\it FIRSST}, \citealt{cooray23}), is a 2m class aperture telescope with spectroscopic capability over the 40-260$\, \umu$m range, again allowing observations to probe below the photometric confusion limit. 

Finally, the {\it SPace Interferometer for Cosmic Evolution} ({\it SPICE}, \citealt{leisawitz23}), is an alternative concept. A cryogenic (4K) spatial-spectral interferometer with two connected 1m dishes with a variable baseline between 6 - 36m, providing an angular resolution of 0.3{\arcsec} at 100$\, \umu$m (0.75{\arcsec} at the SPIRE 250$\, \umu$m band).  {\it SPICE} will be able to image fields of 1$\arcmin \times$ 1$\arcmin$ with the caveat of taking of the order of 24 hours to get decent uv coverage of such a field.

What is clear, is that the SPIRE Dark Field currently represents the deepest image of the far-infrared sky that will not be surpassed until at least the turn of the decade. 

\smallskip

\section{Acknowledgements}
The authors would like to thank J. Krick for discussions and assistance with {\it Spitzer} data in the IRAC Dark Field. The authors acknowledge the contributions from D. Rasimavicius,  C. Aldegunde Manteca and C. Bonacina.  The authors would like to thank the anonymous referee for their comments that improved the quality and content of this paper. This work was funded in part by a UROP grant from Imperial College Physics Department. CP acknowledges support via the RAL Space In House Research programme funded by the Science and Technology Facilities Council of the UK Research and Innovation (award ST/M001083/1).

SPIRE has been developed by a consortium of institutes led by Cardiff Univ. (UK) and including: Univ. Lethbridge (Canada); NAOC (China); CEA, LAM (France); IFSI, Univ. Padua (Italy); IAC (Spain); Stockholm Observatory (Sweden); Imperial College London, RAL, UCL-MSSL, UKATC, Univ. Sussex (UK); and Caltech, JPL, NHSC, Univ. Colorado (USA). This development has been supported by national funding agencies: CSA (Canada); NAOC (China); CEA, CNES, CNRS (France); ASI (Italy);MCINN (Spain); SNSB (Sweden); STFC, UKSA (UK); and NASA (USA). HIPE is a joint development by the Herschel Science Ground Segment Consortium, consisting of ESA, the NASA Herschel Science Center, and the HIFI, PACS and SPIRE consortia. 

For the purpose of open access, the author has applied a Creative Commons Attribution (CC BY) licence (where permitted by UKRI, 'Open Government Licence' or 'Creative Commons Attribution No-derivatives CC BY-ND licence' may be stated instead) to any Author Accepted Manuscript version arising

\section{Data Availability}
All Herschel data is available on the public ESA archive at {\tt{http://archives.esac.esa.int/hsa/whsa/}}. Source catalogues and source counts will be made available on reasonable request.



\bsp 

\label{lastpage}

\end{document}